\documentclass[12pt]{article}
\usepackage[latin1]{inputenc}

\usepackage{amsmath}
\usepackage{amsfonts}
\usepackage{amssymb}
\usepackage{graphicx}
\usepackage{geometry}
\usepackage{amssymb,epsfig,subfigure}
\usepackage{hyperref}
\usepackage[bottom,flushmargin,hang,multiple]{footmisc}
\usepackage{etoolbox}
\makeatletter
\patchcmd\maketitle{\hb@xt@1.8em}{\hbox}{}{}
\makeatother

\makeatletter
\renewcommand\section{\@startsection {section}{1}{\z@}%
                                 {-3.5ex \@plus -1ex \@minus -.2ex}
                                   {2.3ex \@plus.2ex}%
                                   {\normalfont\large\bfseries}}
\renewcommand\subsection{\@startsection{subsection}{2}{\z@}%
                                   {-3.25ex\@plus -1ex \@minus -.2ex}%
                                     {1.5ex \@plus .2ex}%
                                     {\normalfont\bfseries}}
\renewcommand\subsubsection{\@startsection{subsubsection}{3}{\z@}%
                                   {-3.25ex\@plus -1ex \@minus -.2ex}%
                                     {1.5ex \@plus .2ex}%
                                     {\normalfont\itshape}}
\makeatother

\def\pplogo{\vbox{\kern-\headheight\kern -29pt
\halign{##&##\hfil\cr&{\ppnumber}\cr\rule{0pt}{2.5ex}&\ppdate\cr}}}
\makeatletter
\def\ps@firstpage{\ps@empty \def\@oddhead{\hss\pplogo}%
  \let\@evenhead\@oddhead 
}
\def\maketitle{\par
 \begingroup
 \def\thefootnote{\fnsymbol{footnote}}
 \def\@makefnmark{\hbox{$^{\@thefnmark}$\hss}}
 \if@twocolumn
 \twocolumn[\@maketitle]
 \else \newpage
 \global\@topnum\z@ \@maketitle \fi\thispagestyle{firstpage}\@thanks
 \endgroup
 \setcounter{footnote}{0}
 \let\maketitle\relax
 \let\@maketitle\relax
 \gdef\@thanks{}\gdef\@author{}\gdef\@title{}\let\thanks\relax}
\makeatother

\numberwithin{equation}{section}

\begin{document}

\setcounter{page}0
\def\ppnumber{\vbox{\baselineskip14pt
}}
\def\ppdate{\footnotesize{SLAC-PUB-15958, SU-ITP-14/07}} \date{}

\author{Danjie Wenren\footnote{djwenren@stanford.edu}\\
[7mm]
{\normalsize \it Stanford Institute for Theoretical Physics }\\
{\normalsize  \it Department of Physics, Stanford University }\\
{\normalsize \it Stanford, CA 94305, USA}\\
[3mm]
{\normalsize \it Theory Group, SLAC National Accelerator Laboratory}\\
{\normalsize  \it Menlo Park, CA 94025, USA}\\
[3mm]}

\bigskip
\title{\bf  Tilt and Tensor-to-Scalar Ratio in Multifield Monodromy Inflation
\vskip 0.5cm}
\maketitle

%
%
%
%

\begin{abstract}
We study the possible range of the tilt $n_s$ and the tensor-to-scalar ratio $r$ in multifield versions of a class of inflationary models from string theory. We show that $r$ is the same between the single field models and multifield models while $n_s$ is bounded above by the results of single field models. Below its maximum value, $n_s$ depends on the specific distributions of parameters in the model. The general trend is that the wider the distributions are, the smaller $n_s$ is. We show that $n_s$ does not have a rigorous lower bound. It is argued, however, that models predicting arbitrarily small $n_s$ only constitute a small portion of the possible ones and for the vast majority of models, $n_s$ is bounded below by predictions given by models with uniformly distributed parameters.
\end{abstract}
\bigskip
\newpage


\vskip 1cm

\maketitle

\section{Introduction}
An important implementation of inflaton in string theory is the numerous axions arising from 
integrating fluxes around non-trivial cycles on the 
compactification manifold. The mechanism of monodromy arises very naturally in models from string theory \cite{AxionMonodromy:1, AxionMonodromy:2}. 
It can break the discrete periodicity in the axion field range and thus extend the kinetic region, leading to a UV completion of chaotic inflation \cite{chaotic} with some residual features of natural inflation \cite{natural}. 
In models which do not consider this effect, the rolling range of the inflaton is usually sub-Planckian. 
To achieve the number of e-foldings required, one solution is to consider models with numerous inflaton fields \cite{Dimopoulos:2005ac, Liddle:1998jc}. 
In such models, the inflatons can collectively drive the inflation phase and each of them only needs to roll 
a relatively small range.  

However, in a given direction in field space the monodromy effect seems generic (in that avoiding it requires turning off various fluxes and branes), as is a multiplicity of axion fields, so it is interesting to simply consider the two effects together.  We will see this has an interesting effect on the phenomenology, pushing the tilt further toward the central region that is observationally viable relative to the predictions of either single-field monodromy models or N-flation without monodromy.     


The single-field axion monodromy models do not restrict the rolling range of the inflaton to be 
sub-Planckian, though given the normalized power spectrum they do have upper bounds for field ranges due to the requirement that the inflaton potential not exceeding the moduli stabilization potential. It has been shown rather explicitly in \cite{AxionMonodromy:2} that one can construct such models with super-Planckian field range, realizing the mechanism proposed in \cite{AxionMonodromy:1}. Although it was shown in \cite{Gur-Ari:2013sba} that there is some technical difficulty in realizing the moduli stabilization used in the specific models discussed originally in \cite{AxionMonodromy:1}, we expect this mechanism to be solid and more general \cite{flattening}. The specific parameters 
of these models, however, can be very different from each other. For instance, \cite{Gur-Ari:2013sba} shows that at the level of monodromies available twisted tori one can get candidate power law potentials like $V(\phi) \sim \phi^{2/3},~\phi,~\phi^{6/5},~\phi^{4/3},~\phi^{10/7},~\phi^{3/2},~\phi^{2}$, where $\phi$ is the inflaton field, by generalizing the models discussed in \cite{AxionMonodromy:1, AxionMonodromy:2}, and similarly in \cite{flattening}.
These examples as well as more recent works \cite{recentmonodromy}\ suggest that the power of the inflaton field in the potential can take various 
values and thus the space of predictions that such kind of models can make is nontrivial.  

In this work we consider models that combine the two ideas, that is, multifield versions of axion monodromy models. In fact, it is natural to think that 
there can be many inflaton fields in models from string theory, given the large number of 
possible compactification manifolds 
and various ways of wrapping the branes. We will show in secton 
\ref{sec:StrictBound} and \ref{sec:stats} that such models, assuming that they do arise, can give a range of possible predictions for the 
spectral index $n_s$, with an 
upper bound given by the corresponding single-field models. Most of those predictions are within 
the region given by the experimental results \cite{Plack2013:I, Ade:2014xna}. 

In principle different inflaton fields in a multifield model can have different powers in their potential $V(\phi)\sim \phi^p$, but since we are interested in the dependence of the predictions of the 
model on the power $p$, we assume for simplicity that for a given model, it is the same for all 
the fields. So for a model with $N$ inflaton fields, the potential of inflaton $i$ is taken to be
$V_i = \mu_i \phi_i^p$. Note that the coefficients $\mu_i$ are 
dimensional quantities, with $[\mu_i] = 4-p$. We take the Lagrangian for the inflatons to be  
\begin{equation}
  \mathcal{L} = \sqrt{-g}\sum_i\left(-\frac{1}{2}(\partial\phi_i)^2 - V_i\right), 
\end{equation}
where we have assumed that the inflatons only couple indirectly via gravity. The equations of motion are 
\begin{align}
  \label{eom:a} &H^2 = \left(\frac{\dot{a}}{a}\right)^2 = \frac{1}{3}\sum_i\left(\frac{1}{2}\dot{\phi_i}^2 + V_i\right) \\%
  \label{eom:phi} &\ddot{\phi_i} + 3H\dot{\phi}_i + V'_i = 0.
\end{align}
For notational simplicity we will work in Planck units. As noted in \cite{LargeMassRollFaster:1, NumericNflation}, 
   this set of equations of motion leads to 
solutions with larger $\mu_i$ fields rolling faster than those with smaller $\mu_i$, and thus the larger $\mu_i$ fields 
exiting the inflation phase earlier. In fact if we view $\vec{\phi} = (\phi_1, \cdots, \phi_N)$ as a vector in the field space, then in the slow roll limit, (\ref{eom:phi}) can be written as $3H \dot{\vec{\phi}} \approx -\nabla_{\phi}V$, showing that inflation follows the steepest trajectory of the potential in the field space. This has a strong implication for the dynamics towards the end of the inflation phase. However in this work we consider the spectral index and the tensor-to-scalar ratio predicted by the multifield models, which depend on $\mu_i$ and the initial values of $\phi_i$. Therefore we can assume that the system is still in the slow-roll region.

In multifield models, the two conditions from the number of e-foldings and the normalization scale of scalar curvature perturbation cannot completely fix the field values or the coefficients $\mu_i$'s. As a result the prediction of $n_s$ depends on other conditions. We will show that given the power of the potential $p$, the maximum possible value for $n_s$ is given by the single field models. On the other hand, we will also show that the tensor-to-scalar ratio $r$ can be determined by the number of e-foldings in the multifield models considered here.

Below the maximum value of $n_s$, multifield models can also predict a range of possible values for it. The exact prediction, however, will depend on the configuration of the inflaton initial values and the 
coefficients $\mu_i$. There might be some top-down mechanism to fix them, but 
without the knowledge of such mechanisms, one approach to the problem is to 
allow those values to be random 
and ask what predictions can be made given the two conditions of number of e-foldings and normalization scale of scalar curvature perturbation. When the number of 
fields are large, the Central Limit Theorem indicates that the observables can be determined by the distributions of $\phi$'s and $\mu$'s to very good precision, not depending on the particular values of $\phi_i$'s or $\mu_i$'s. 

In a multifield model, the number of e-foldings achieved by slow roll is 
\begin{eqnarray}
  N_e &=& \int^{t_f}_{t_i}H dt = \int^{t_f}_{t_i}\frac{\sum_i\left(\frac{1}{2}\dot{\phi}^2_i + V_i\right)}{3H}dt 
          \approx \sum_i\int^{t_f}_{t_i}\frac{V_i}{3H}dt \nonumber \\%
      &\approx& -\sum_i\int^{t_f}_{t_i}\frac{V_i}{V'_i}\dot{\phi}_i dt = \sum_i\int^{\phi_{i, \rm{init}}}_{\phi_{i, \rm{final}}}\frac{V_i}{V'_i}d\phi_i.
      \label{EFolding:1}
\end{eqnarray}
In the first approximation we used $\dot{\phi}_i^2 \ll V_i$ and in the second one we used $3H\dot{\phi}_i\approx -V'_i$. 
Evaluate this with $V_i=\mu_i\phi^p_i$, then
\begin{equation}
  N_e \approx \frac{1}{2p}\sum_i\phi^2_i,
  \label{condition:EFolding}
\end{equation}
where we have left out the contribution from the field values at the end of inflation. 

Another condition is from scalar curvature perturbation, which in the slow roll limit can be expressed as (\cite{AnalyticSpec, SpecNFlation})
\begin{equation}
  \mathcal{P}_R = \left(\frac{H}{2\pi}\right)^2\frac{\partial N_e}{\partial \phi_i}\frac{\partial N_e}{\partial \phi_j}\delta_{ij} 
    \approx \frac{V}{12\pi^2}\sum_i\left(\frac{V_i}{V'_i}\right)^2 = \frac{V}{12\pi^2 p^2}\sum_i\phi_i^2,
\end{equation}
where the approximation is due to the slow roll limit. Plugging in the condition of $N_e = (1/2p)\sum_i \phi^2_i$, we get 
\begin{equation}
  \mathcal{P}_R = \frac{V}{6\pi^2 p}N_e.
\end{equation}
This gives the other condition
\begin{equation}
  \label{condition:PR}
  \sum_i \mu_i \phi^p_i = \frac{6\pi^2 p}{N_e}\mathcal{P}_R.
\end{equation}

The spectral index of scalar curvature perturbation is (\cite{AnalyticSpec, SpecNFlation}) 
\begin{eqnarray}
  n_s - 1 &=& 2\frac{\dot H}{H^2} - 2\frac{N_{e,i}\left[(\dot{\phi_i}\dot{\phi_j})/H^2 - (V'_i)'_j/V\right]N_{e,j}}{\delta_{ij}N_{e,i}N_{e,j}} \nonumber \\%
        &\approx& -\frac{1}{V^2}\sum_i(V'_i)^2 -\frac{2}{\sum_i(V_i/V'_i)^2} + \frac{2}{V}\left[\frac{\sum_i V''_i(V_i/V'_i)^2}{\sum_i(V_i/V'_i)^2}\right],
\end{eqnarray}
where ``$N_{e,i}$'' means the derivative of $N_e$ with respect to $\phi_i$. The approximation is again due to the slow roll limit. 
Now using the form of the potential and the two conditions on $\sum_i\phi^2_i$ and $\sum_i \mu_i\phi^p_i$, $n_s -1$ becomes 
\begin{equation}
  \label{ns:expression}
  n_s - 1 = -\frac{1}{N_e} - \frac{N_e^2}{36\pi^4 \mathcal{P}_R^2}\sum_i\left(\mu_i\phi_i^{p-1}\right)^2 
\end{equation}

The tensor perturbation is (\cite{NumericNflation})
\begin{equation}
	\mathcal{P}_g = \frac{2H^2}{\pi^2},
\end{equation}
therefore the tensor-to-scalar ratio can be written as 
\begin{eqnarray}
  r &=& \frac{\mathcal{P}_g}{\mathcal{P}_R}\approx \frac{8}{\sum_{i}\left(V_i/V'_i\right)^2} \nonumber \\%
    &=& \frac{8p^2}{\sum_i\phi_i^2} = \frac{4p}{N_e}.
\end{eqnarray}
This shows that the tensor-to-scalar ratio is completely fixed by the condition from 
the number of e-foldings, independent of the number of the fields 
or the distributions of $\mu$ or $\phi$, as first noted by \cite{LythBound}. 
Therefore in the following we will mainly focus on the estimation of the 
spectral index.

In section \ref{sec:StrictBound} we show that $n_s$ is strictly bounded above by the results of single-field models. We also discuss the possible unphysical configurations of $\mu_i$'s and $\phi_i$'s that would lead to arbitrarily small $n_s$. In section \ref{sec:stats} we show that the general dependence of $n_s$ on the distributions of $\mu_i$'s and $\phi_i$'s is that the wider the distributions are, the smaller $n_s$ is. It is also argued that for the vast majority of possible distributions, $n_s$ is within a range bounded by (\ref{ns:max}) and (\ref{ns:min}).

\section{Strict bound of $n_s$}\label{sec:StrictBound}
The unknown part in (\ref{ns:expression}) is $\sum_i\mu^2_i \phi_i^{2p-2}$. In this section 
we will solve for its minimum value and thus the maximum value for $n_s$, subject to the 
conditions (\ref{condition:EFolding}) and (\ref{condition:PR}). We will also show that given only the two conditions, there is no lower bound for $n_s$. 

To find the upper bound on $n_s$, we 
use the Lagrange multiplier method. Let $g$ be
\begin{equation}
  g ~=~ \sum_i\mu^2_i \phi_i^{2p-2} - \alpha\left(\sum_i \phi_i^2 - 2pN_e\right) - 
    \beta\left(\sum_i \mu_i \phi_i^p - \frac{6\pi^2 p}{N_e}\mathcal{P}_R\right),
\end{equation}
where the two expressions inside the parentheses will be set to zero by the equations of motion of the two Lagrange multipliers $\alpha$ and $\beta$ and thus the conditions (\ref{condition:EFolding}) and (\ref{condition:PR}) are satisfied. 
At the extremal values of $g$, $\mu_i$ and $\phi_i$ satisfy the following relations
\begin{eqnarray}
  \label{extrm:a} \sum_i \phi_i^2 - 2pN_e = 0, \qquad
  \label{extrm:b} \sum_i \mu_i \phi_i^p - \frac{6\pi^2 p}{N_e}\mathcal{P}_R = 0 \\%
  \label{extrm:mu} 2\mu_i \phi_i^{2p-2} - \beta\phi_i^p = 0, \qquad
  \label{extrm:phi} (2p-2)\mu_i^2\phi_i^{2p-3} - 2 \alpha \phi_i - \beta\mu_i p \phi_i^{p-1} = 0,
\end{eqnarray}
which are obtained by taking the derivative of $g$ with respect to $\alpha$, $\beta$, $\mu_i$ and $\phi_i$, respectively. 
Since $\phi_i > 0$, the solution to the set of equations above is
\begin{equation}
  \beta = \frac{6\pi^2}{N_e^2}\mathcal{P}_R,~
  \alpha = -\frac{1}{4}\beta^2,~
  \label{MuiInPhii} \mu_i = \frac{3\pi^2\mathcal{P}_R}{N_e^2}\phi_i^{2-p},
\end{equation}
and $\phi_i$ can be any value as long as $\sum_i\phi_i^2 = 2pN_e$.

To show that the above solution corresponds to the upper bound on $n_s$, let 
\begin{equation}
	\mu_i = (3\pi^2\mathcal{P}_R/N_e^2)\phi_i^{2-p} + \delta\mu_i,
\end{equation}
then the 
condition (\ref{condition:PR}) yields
\begin{equation}
  \sum_i(\delta\mu_i)\phi_i^p = 0.
\end{equation}
The spectral index becomes
\begin{equation}
  \label{ns:transformed}
  n_s = 1 -\frac{1}{N_e}\left(1+\frac{p}{2}\right) - \frac{N_e^2}{36\pi^4\mathcal{P}_R^2}
    \sum_i\left(\delta\mu_i\right)^2\phi_i^{2p-2}.
\end{equation}
Since the last term is always non-positive, $n_s$ has a maximum value 
\begin{equation}
  n^{\max}_s = 1 - \frac{1}{N_e}\left(1+\frac{p}{2}\right),
\end{equation}
which is exactly the result given by single-field models\footnote{One can also show the upper bound on $n_s$ by using the Cauchy-Schwarz inequality, that is, $(\sum_i\mu_i^2 \phi_i^{2p-2})(\sum_i\phi_i^2) \ge (\sum_i\mu_i\phi_i^p)^2$. The equality holds iff $\mu_i^2\phi_i^{2p-2}/\phi_i^2$ is the same constant for all $i$. This condition is always satisfied by the single-field models since there is only one such ratio. Therefore the maximum value corresponds to that given by the single field models.}. (\ref{ns:transformed}) also shows that 
the greater $\mu_i$ deviates from $(3\pi^2\mathcal{P}_R/N_e^2)\phi_i^{2-p}$, the smaller 
$n_s$ is.

On the other hand, one might find that $n_s$ does not have a lower bound when only conditions (\ref{condition:EFolding}) and (\ref{condition:PR}) are imposed, that is, without insisting on the slow-roll condition. For example, in models with $p<1$, there might be initial conditions where some $\phi_i$ are aggregated near some very small number. The conditions can still be satisfied by making some other $\phi_i$ large but the corresponding $\mu_i^2 \phi_i^{2p-2}$ can be made very large and thus $n_s$ very small. Moreover, even when $p>1$ one can still make the $n_s$ arbitrarily small by distributing $\phi_i$ and $\mu_i$ wide enough. For instance one can consider the situation where one $\mu_i$ takes the maximum possible value $\mu_m$ and the other $(N-1)$ $\mu_i$'s are zero. Assume that all the $\phi_i$ take the same value $\phi = (2pN_e/N)^{1/2}$. Then the condition (\ref{condition:PR}) gives $\mu_m = (6\pi^2 p \mathcal{P}_R)/(N_e \phi^p)$. Then 
\begin{equation}
	\sum_i \mu_i^2 \phi_i^{2p-2} = \mu_m^2 \phi^{2p-2} = \frac{18\pi^4 p\mathcal{P}_R^2}{N_e^3}N,
\end{equation}
which grows linearly with the number of inflatons, and thus $n_s$ can be arbitrarily small when the number of inflatons is large. 

We should emphasize that the situation in the previous paragraph is not physical as it violates the slow roll condition and the validity of the power law potential approximation of the underlying potential from string theory. The slow roll parameter in the multifield models considered here is 
\begin{equation}
	\label{slowroll}
	\epsilon = -\frac{\dot{H}}{H^2} \approx \frac{N_e^2}{72 \pi^4 \mathcal{P}_R^2}\sum_i \mu_i^2 \phi_i^{2p-2},
\end{equation}
where the approximation is due to the slow roll condition. Therefore the requirement of $\epsilon$ being small shows that the extreme example considered in the previous paragraph is physically not allowed. In the meantime, the power law potential is valid in the region where the field values $\phi_i$ are much bigger than some nonzero scales in the detailed model building \cite{AxionMonodromy:1, AxionMonodromy:2, Gur-Ari:2013sba}. When most of the field values become very small, which normally only happens near the end of inflation, the potential should be of another form. Another problem with the special situation above is that the fields with $\phi_i$ and $\mu_i$ very close to 0 (in Planck units) do not actually contribute to the inflation and should not be considered as inflatons. In fact, we dropped the terms from $\phi_{i,\mathrm{final}}$ in (\ref{EFolding:1}) using the assumption that there is a large separation between $\phi_{i,\mathrm{init}}$ and $\phi_{i,\mathrm{final}}$, which is valid when $\phi_i$ contribute to inflation. However, when the field initial values are very close to 0, $\phi_{i,\mathrm{init}}$ and $\phi_{i,\mathrm{final}}$ are roughly the same and those $\phi_i$'s do not contribute to inflation.

\section{Statistical analysis}\label{sec:stats}
The previous section shows that single field models give the maximum spectral index among models with the same $p$ and that the maximum value can also be reached given that $\mu_i$ is correlated with $\phi_i$ as in (\ref{MuiInPhii}). In this section, we explore the possible values of the spectral index away from the maximum value. As mentioned early, we assume that the number of inflatons is large and the parameters $\mu_i$ and $\phi_i$ are random. 

We will also assume that the distributions of $\mu_i$ and $\phi_i$ are independent in the following. That might not be the case and in fact might have a large effect on the result. As we have shown in the last section, if $\mu$ is correlated with $\phi$ as given by (\ref{MuiInPhii}), $n_s$ is always at its maximum regardless of the distribution of the distribution of $\phi$. However in most models of chaotic inflation, we do not expect such a correlation between the form of the inflaton potential and the initial values of the inflaton fields. Hence the assumption is a good working approximation. 

One of the most important differences between the single field models and multifield models is that in the latter, the field values and coefficients can be spread over a range while the former can be regarded as the limit where the distributions are $\delta$-functions. This means that in multifield models, the observable predictions may depend on how wide the initial field values and coefficients are spread. There might also be other factors in determining the predictions, for instance, the number of peaks in the distribution or the location of the peaks. 

There have been works on parameter distributions of some models. For example, \cite{SpecNFlation} uses results from random matrix theory to explore the hypothesis that the masses of the axions in N-flation models follow the Marcenko-Pastur distribution. Here we do not consider any particular top-down construction of distributions, rather we investigate what predictions there can be amongst a large class of distributions. We will mainly show the effect of peak width of the distributions. The specific form of the distribution is not expected to have a qualitative effect since any class of distributions over a finite range will interpolate between the $\delta$-fuction case and the uniform distribution case when the widths of peaks go from small to large.

Since the inflaton fields and the coefficients both have maximum values, determined by the specific model building, the distributions $f(x)$ they follow should only be non-zero in $(0,x_m]$ for some positive $x_m$. 
The lower bound is set to be 0 since the minimum value of $\phi$ and $\mu$ can be made well below the maximum value. It is equivalent to the assumption we used earlier that $\phi_{i, \mathrm{final}}$ can be neglected in (\ref{EFolding:1}). It also simplifies the problem if we rescale $\phi$ and $\mu$ as $\phi=\phi_m \hat{\phi}$ and $\mu = \mu_m \hat{\mu}$, respectively, such that $\hat{\phi}$ and $\hat{\mu}$ distribute between $0$ and $1$. 

When the number of inflatons $N$ is large, the Central Limit Theorem indicates that the condition (\ref{condition:EFolding}) and (\ref{condition:PR}) can be approximated by 
\begin{equation}
	\label{condition:combined}
	\langle\phi^2\rangle=\phi_m^2\langle\hat{\phi}^2\rangle = A,\quad
	\langle\mu\phi^p\rangle = \mu_m\phi_m^p\langle\hat{\mu}\rangle\langle\hat{\phi}^p\rangle = B,
\end{equation} 
where we have let $A=\frac{2pN_e}{N}$ and $B = \frac{6\pi^2 p\mathcal{P}_R}{N_e N}$. This gives
\begin{equation}
	\phi_m = \frac{A^{1/2}}{\langle\hat{\phi}^2\rangle^{1/2}}, \quad
	\mu_m = \frac{B}{A^{p/2}}\frac{\langle\hat{\phi}^2\rangle^{p/2}}{\langle\hat{\phi}^p\rangle}\frac{1}{\langle\hat{\mu}\rangle}.
\end{equation}
 Note that we used the assumption that $\mu$ and $\phi$ are independent in eq. (\ref{condition:combined}). 

The spectral index (\ref{ns:expression}) can also be written as 
\begin{equation}
	\label{ns:expression:apprx}
	n_s - 1 = -\frac{1}{N_e} - \frac{N_e^2}{36\pi^4\mathcal{P}_R^2}N\langle\mu^2\phi^{2p-2}\rangle.
\end{equation}
The unknown quantity in (\ref{ns:expression:apprx}) is
\begin{equation}
	\langle\mu^2\phi^{2p-2}\rangle = \frac{B^2}{A}\frac{\langle\hat{\mu}^2\rangle}{\langle\hat{\mu}\rangle^2}\frac{\langle\hat{\phi}^2\rangle\langle\hat{\phi}^{2p-2}\rangle}{\langle\hat{\phi}^p\rangle^2}.
\end{equation}
Using this and plugging in the expressions of $A$ and $B$, we have 
\begin{equation}
	\label{ns:expression:apprx:2}
	n_s = 1 - \frac{1}{N_e}\left(1+\frac{p}{2} \frac{\langle\hat{\mu}^2\rangle}{\langle\hat{\mu}\rangle^2}\frac{\langle\hat{\phi}^2\rangle\langle\hat{\phi}^{2p-2}\rangle}{\langle\hat{\phi}^p\rangle^2}\right).
\end{equation}
Note that (\ref{ns:expression:apprx:2}) shows when the number of field gets large enough such that the prediction only depends on the distributions of $\hat{\mu}$ and $\hat{\phi}$, the dependence of the number of fields drops out.

Similar to the discussion at the end of section \ref{sec:StrictBound}, there are distributions of $\hat{\mu}$ and $\hat{\phi}$ with which $n_s$ does not have a lower bound. One example would be the distribution where $\hat{\mu} = 1$ with probability $c$ and otherwise 0. Then $\langle\hat{\mu}\rangle=\langle\hat{\mu}^2\rangle = c$, so 
\begin{equation}
	\frac{\langle\hat{\mu}^2\rangle}{\langle\hat{\mu}\rangle^2} = \frac{1}{c},
\end{equation}
which can be made arbitrarily large by decreasing $c$. However, this type of distributions are not physical for the reasons mentioned at the end of section \ref{sec:StrictBound}. In the following we will consider a large set of distributions, in which those unphysical ones are included for the purpose of showing that they only constitute a very small subset of all the possible distributions.

We will show that the major effect of the distributions on the spectral index is from the width of the peaks. Except for a very small part of the distribution space, which corresponds to the examples mentioned in the previous paragraph, the spectral index are within a well bounded region. We will also show that the number of peaks does not affect the result qualitatively.

\subsection{Peak at the center}\label{subsec:3_1}
\begin{figure}
  \centering
  \includegraphics[width=.7\textwidth]{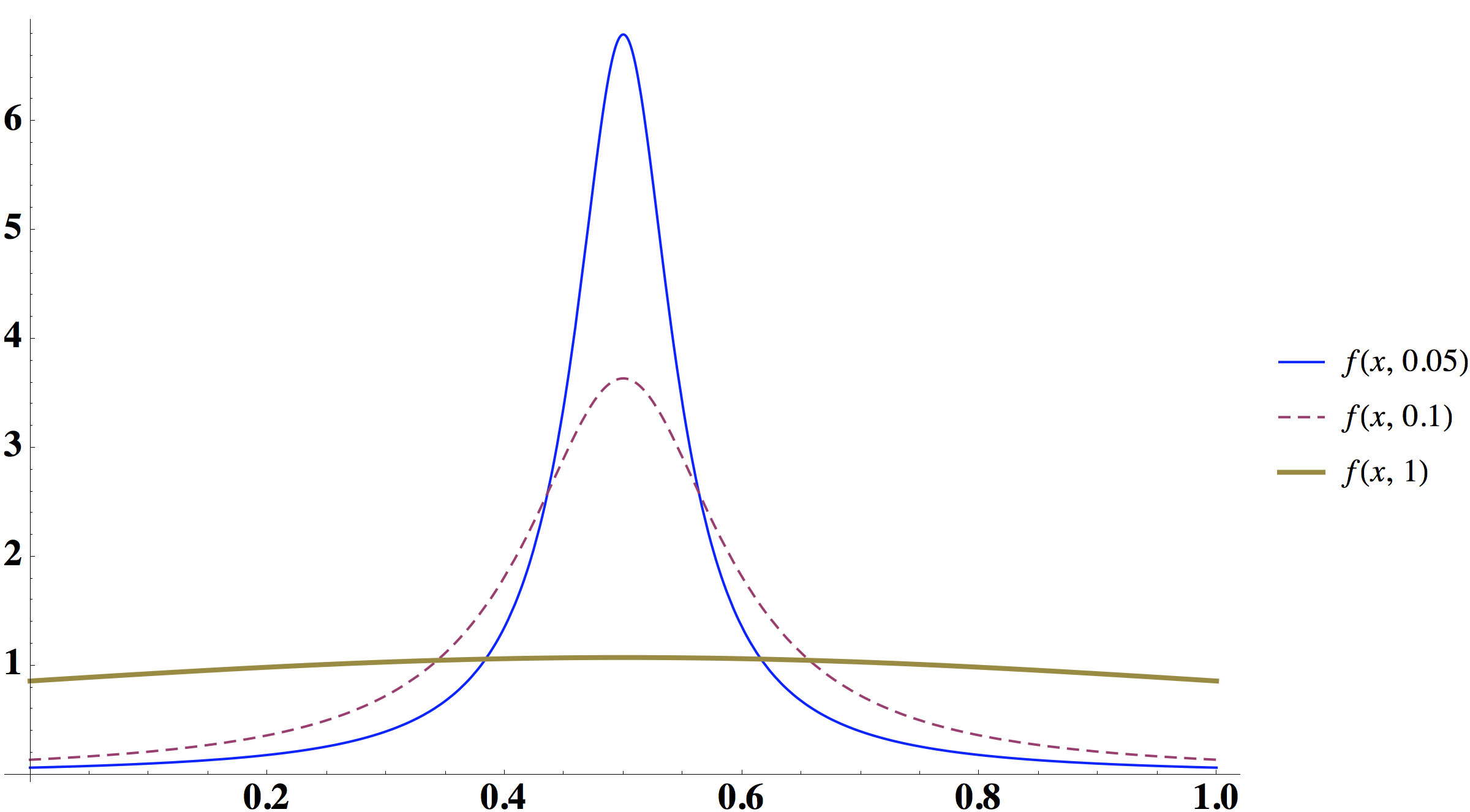} 
  \caption{The distribution function used with different parameters. 
    As expected, the distribution becomes wider as $\epsilon$ gets larger, ultimately approaching a uniform distribution.}
  \label{dist:fx}
\end{figure}

In this subsection we assume that the peak of the distribution is at $1/2$ and 
show the effect of the width of the peak. One convenient distribution is
\begin{equation} 
  \label{dist:center}
  f(x; \epsilon) = 
  \begin{cases}
    \frac{1}{2\arctan(1/2\epsilon)}\frac{\epsilon}{\left(x-1/2\right)^2 + 
        \epsilon^2}, & 0\le x \le 1 \\%
    0, & \rm{otherwise}
  \end{cases}
  .
\end{equation}
The parameter $\epsilon$ sets the width of the peak. When $\epsilon$ goes to zero, this becomes $\delta(x-1/2)$; when it goes to 
$\infty$, this becomes a uniform distribution between $0$ and $1$. 
When $0 < \epsilon < \infty$, the shape of the distribution interpolates between those two distributions, as shown in Fig \ref{dist:fx}. Note that the exact form of the distribution is not crucial in the discussion here since all we care is the features of the distribution, such as the peak width, location, number, etc. 
We assume that $\hat{\phi}$ and $\hat{\mu}$ follow $f(\hat{\phi};\epsilon_{\phi})$ and $f(\hat{\mu};\epsilon_{\mu})$, respectively. 

\begin{figure}
  	\centering
  	\begin{minipage}{0.42\textwidth}
  		\centering
  		\includegraphics[width=\textwidth]{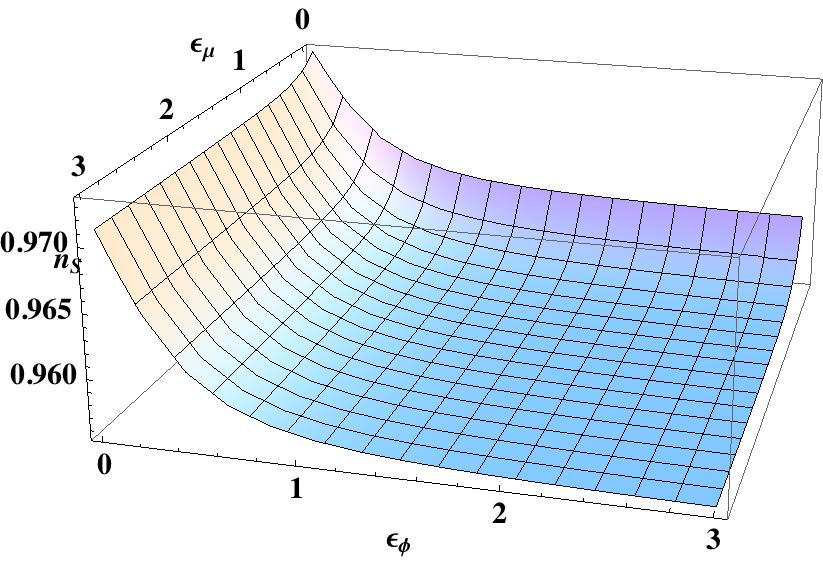}
  	\end{minipage}
  	\begin{minipage}{0.51\textwidth}
  		\centering
  		\includegraphics[width=\textwidth]{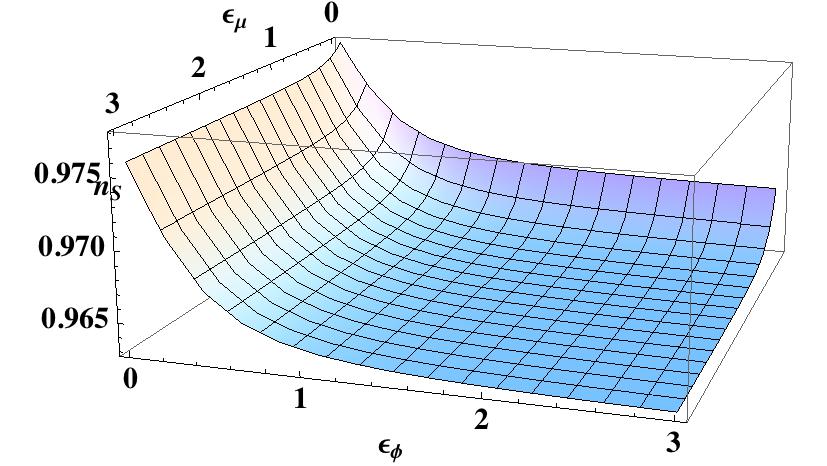}
  	\end{minipage}
  	\caption{$n_s$ of models with $p=2/3$. The number of e-folding is 50 for the first plot and 60 for the second one. The bounded regions given by (\ref{ns:min}) and (\ref{ns:max}) are $[0.955, 0.973]$ and $[0.963, 0.978]$ for the first and second plot, respectively. Note that here $\epsilon_{\mu}$ and $\epsilon_{\phi}$ are both dimensionless since they are the width parameters for the rescaled $\hat{\mu}$ and $\hat{\phi}$.}
  	\label{fig:ep_em_2/3}
\end{figure}

One can use the distribution (\ref{dist:center}) to compute the expectation values in (\ref{ns:expression:apprx}). Motivated by $2/3$ being on the edge of the possible values of $p$ in the models considered in \cite{AxionMonodromy:1, AxionMonodromy:2, Gur-Ari:2013sba} we show the result for this case in Fig \ref{fig:ep_em_2/3}. As we can see, the spectral index approaches the single field value when both the distributions of $\hat{\mu}$ and $\hat{\phi}$ have width $0$ and monotonically approaches the uniform distribution value when the widths become large. Similar calculation can be done for other models. This leads to the result that for the models with one central peak, the prediction for $n_s$ will be between 
\begin{equation}
  \label{ns:max}
  n_s = 1 - \frac{1}{N_e}\left(1+\frac{p}{2}\right),
\end{equation}
and 
\begin{equation}
  \label{ns:min}
  n_s = 1 - \frac{1}{N_e}\left(1+\frac{2p(p+1)^2}{9(2p-1)}\right),
\end{equation}
where (\ref{ns:min}) can be obtained from models with uniformly distributed $\phi_i$'s and $\mu_i$'s. The denominator of (\ref{ns:min}) also shows that models with $p \le 1/2$ is not allowed in the setting here. This is due to the fact that the unknown part in (\ref{ns:expression}) depends on $\phi_i^{2p-2}$. When $p < 1/2$, the power is less than $-1$ and the result strongly depends on the small values of $\phi_i$'s. In that case, the lower bound of the distribution becomes important and cannot be taken as 0. We do not, however, consider those cases in this work.

\subsection{Location of the peak}
\begin{figure}
	\centering
	\begin{minipage}{0.47\textwidth}
		\centering
		\includegraphics[width=\textwidth]{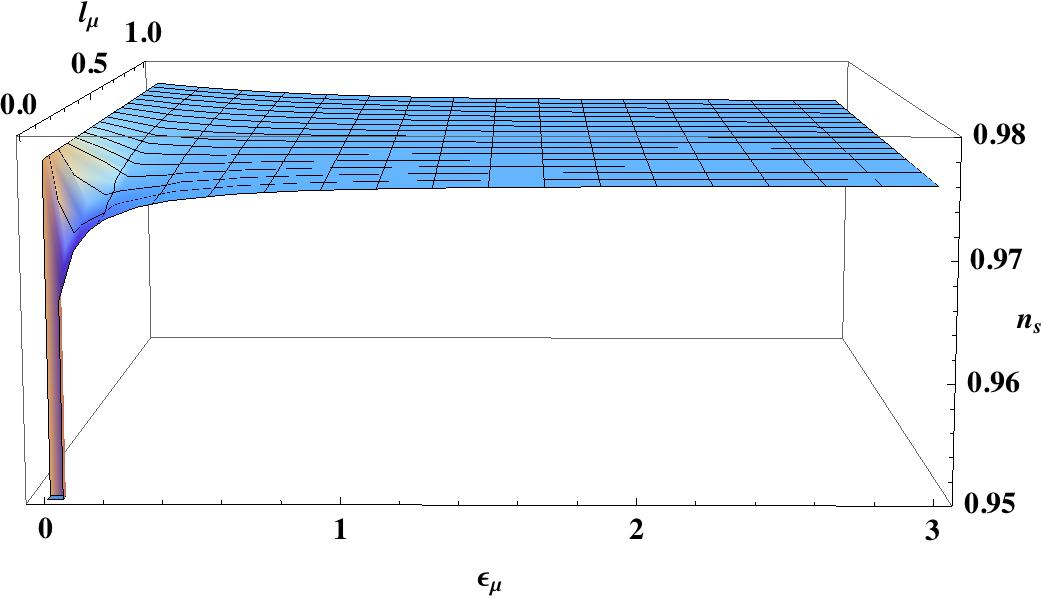}
	\end{minipage}
	\begin{minipage}{0.47\textwidth}
		\centering
		\includegraphics[width=\textwidth]{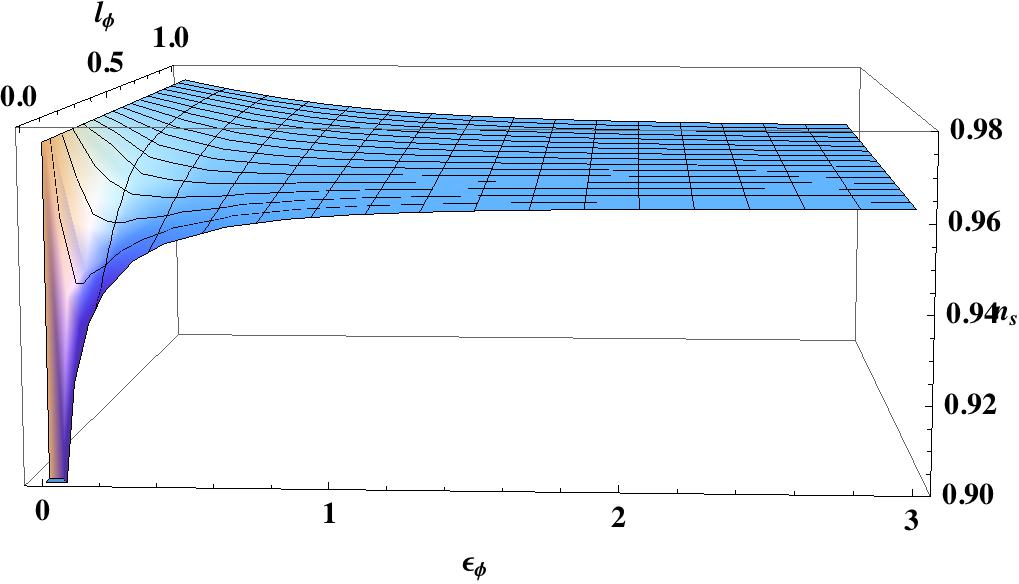}	
	\end{minipage}
	\caption{$n_s$ of models with $p=2/3$ and 60 e-foldings. The bounded region by (\ref{ns:min}) and (\ref{ns:max}) is $[0.963, 0.978]$. Here $l_{\mu}, l_{\phi}\in (0, 1]$ are the locations of the peaks of $\hat{\mu}$ and $\hat{\phi}$, respectively. In the first plot, $\epsilon_{\phi} = 10^{-5}$ and the peak of the $\hat{\phi}$ distribution is at the center. In the second plot, $\epsilon_{\mu} = 1$ and peak of $\hat{\mu}$ is also at the center. Except for the lower left corners, which are argued to violate various conditions at the end of section \ref{sec:StrictBound}, the prediction for $n_s$ is still within the region $[0.963, 0.978]$. }
	\label{fig_wid_loc_2_3}
\end{figure}

The distribution (\ref{dist:center}) can be slightly modified to accommodate for the situation where the peak is not at the center
\begin{equation}
	\label{dist:non-center}
	f(x;\epsilon,l) =
		\begin{cases}
			\frac{1}{\arctan[(1 - l)/\epsilon] + \arctan[l/\epsilon]}\frac{\epsilon}{(x-l)^2 + \epsilon^2}, & 0 \le x \le 1, \\%
			0, & \mathrm{otherwise},
		\end{cases}
\end{equation}
where $0 \le l \le 1$ is the location of the peak. One can again use this distribution to compute the expectation values in (\ref{ns:expression:apprx}). Fig \ref{fig_wid_loc_2_3} shows the effect of the location and width of the peak. One can see that the location of the peak does not have a qualititive effect on the spectral index, except at the lower left corners, which correspond to the unphysical situations mentioned above where essentially all the $\hat{\mu_i}$ and $\hat{\phi_i}$ accumulate near 0. Therefore for the vast majority of the distributions, we can still trust the bounds given in subsection \ref{subsec:3_1}.

\subsection{Multiple peaks}
\begin{figure}
	\centering
	\begin{minipage}{0.47\textwidth}
		\centering
		\includegraphics[width=\textwidth]{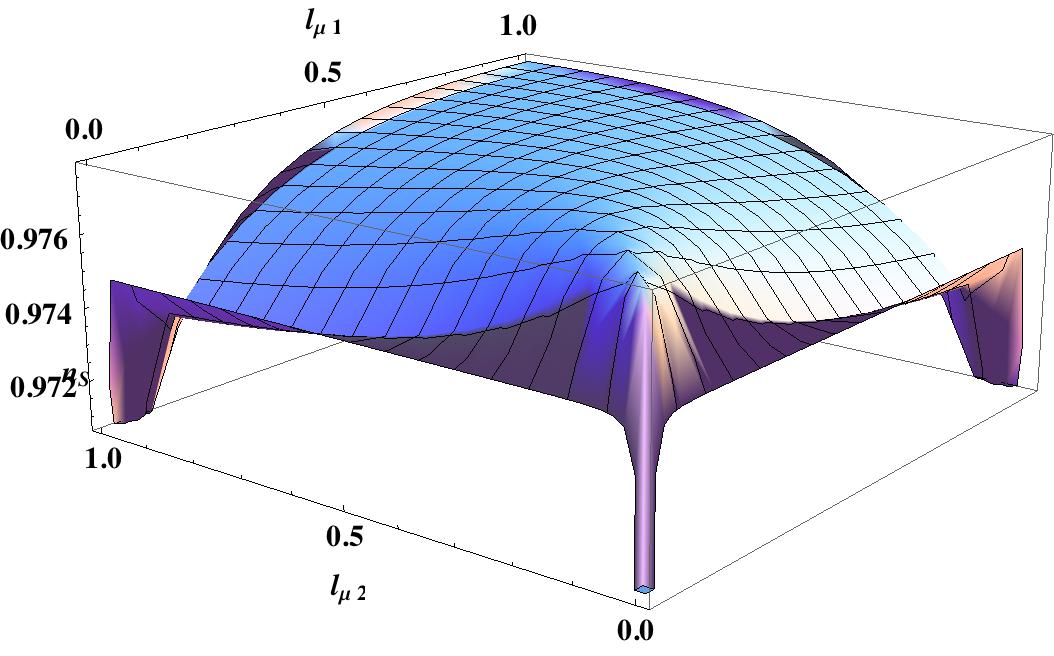}
	\end{minipage}
	\begin{minipage}{0.47\textwidth}
		\centering
		\includegraphics[width=\textwidth]{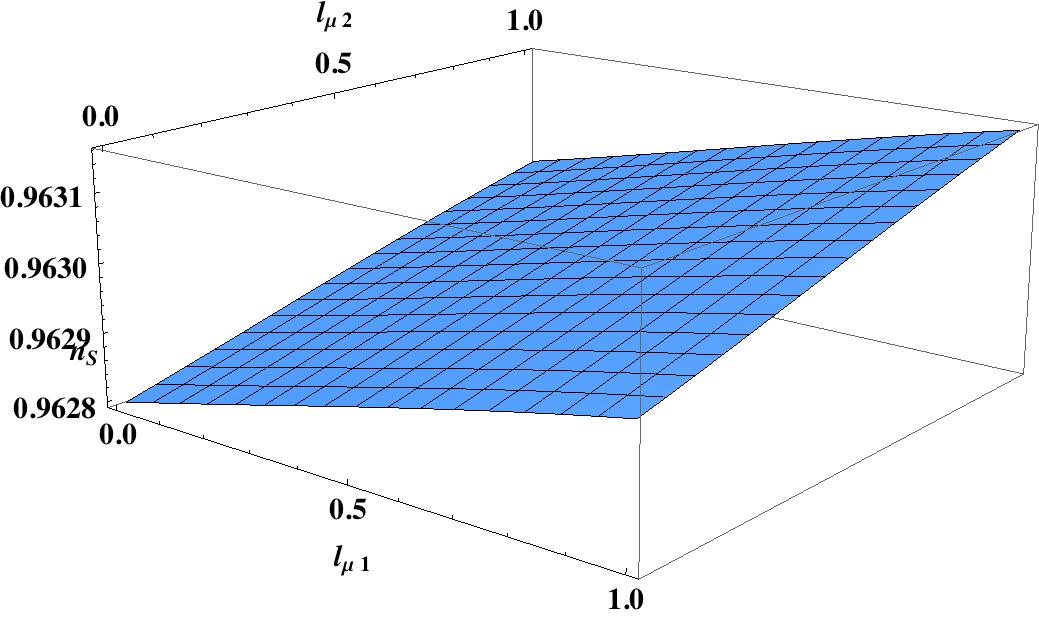}
	\end{minipage}
	\caption{The effects of changing the locations of the two peaks of the distribution of $\hat{\mu}$ in models with $p=2/3$ and $N_e = 60$. $l_{\mu 1}$ and $l_{\mu 2}$ are the locations of the two peaks. In the first plot, all the peaks of $\hat{\mu}$ and $\hat{\phi}$ are very sharp ($\epsilon_{\hat{\mu}},~\epsilon_{\hat{\phi}}\sim 10^{-4}$). In the second plot, the peaks are very wide and the distributions are almost uniform. In both plots we used a distribution of $\hat{\phi}$ with a central peak.}
	\label{fig:multi:loc_mu_2_3}
\end{figure}

\begin{figure}
	\centering
	\begin{minipage}{0.47\textwidth}
		\centering
		\includegraphics[width=\textwidth]{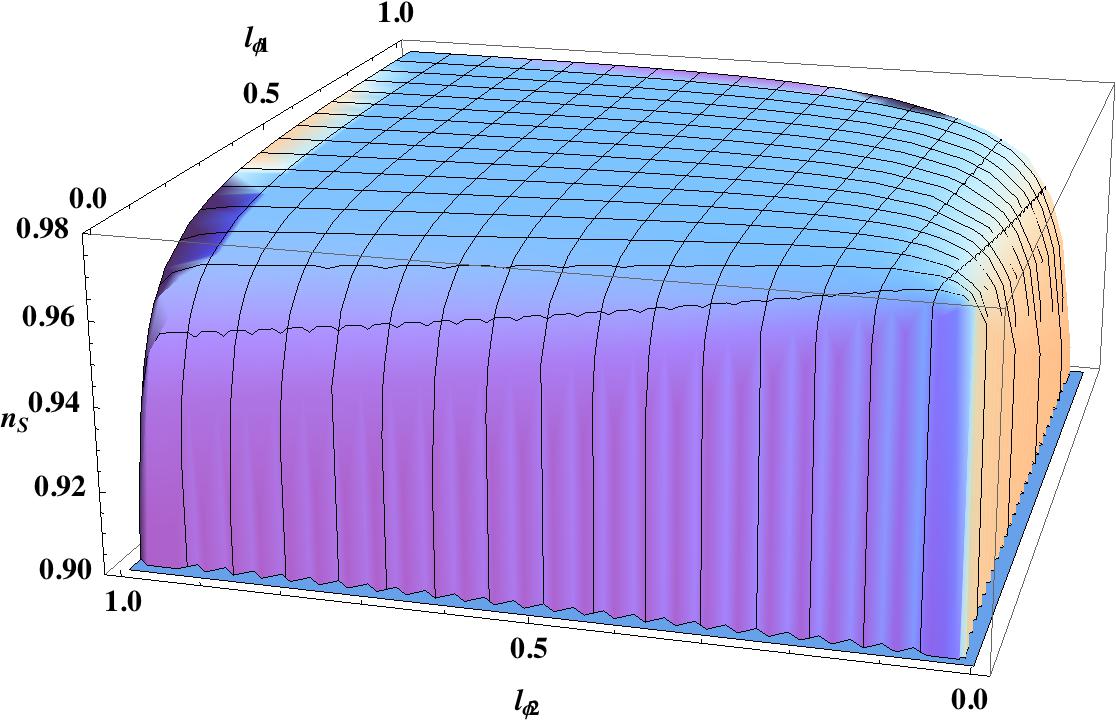}
	\end{minipage}
	\begin{minipage}{0.47\textwidth}
		\centering
		\includegraphics[width=\textwidth]{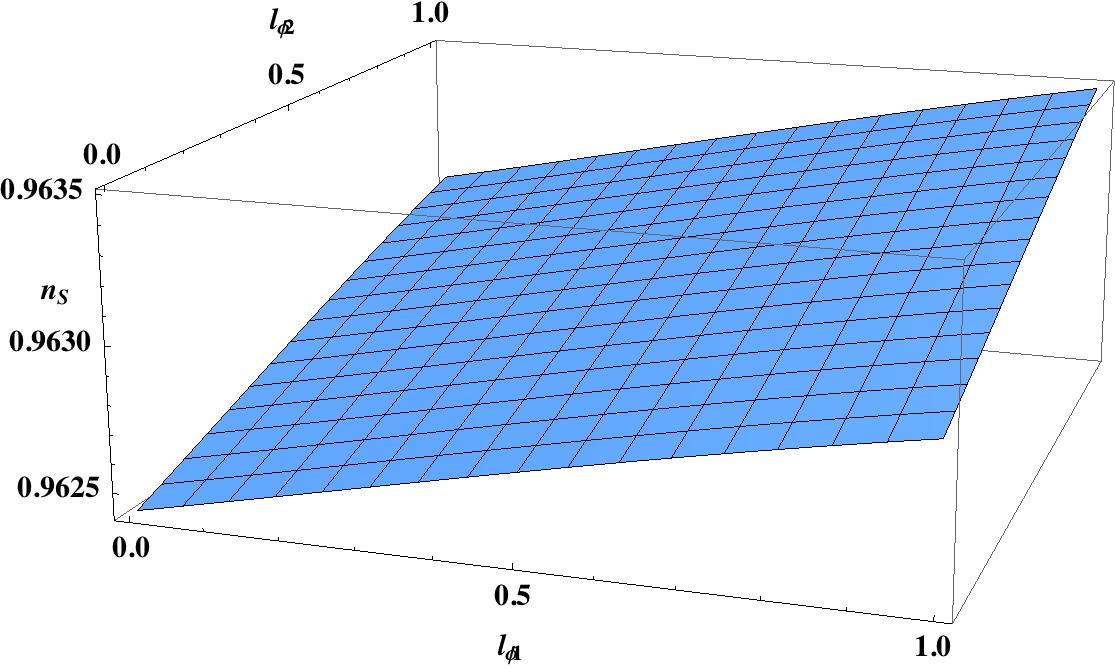}		
	\end{minipage}
	\caption{The effects of changing the locations of the two peaks of the distribution of $\hat{\phi}$ in models with $p=2/3$ and $N_e = 60$. $l_{\phi 1}$ and $l_{\phi 2}$ are the locations of the two peaks. In the first plot, the peaks are very sharp and in the second plot they are very flat. In both plots, the distribution of $\hat{\mu}$ only has a central peak.}
	\label{fig:multi:loc_phi_2_3}
\end{figure}

One can also study the multi-peak situations by using two distributions of (\ref{dist:non-center}), that is,
\begin{equation}
	f(x;\epsilon_1,l_1,\epsilon_2,l_2) = 
		\begin{cases}
			\frac{1}{\arctan[(1 - l_1)/\epsilon_1] + \arctan[l_1/\epsilon_1] + 
 (1 \to 2)}\left(\frac{\epsilon_1}{(x-l_1)^2 + \epsilon_1^2} + (1\to 2) \right), & 0 \le x \le 1, \\%
 			0, & \mathrm{otherwise},
		\end{cases}
\end{equation}
where for simplicity we only consider the situation with two peaks. Fig \ref{fig:multi:loc_mu_2_3} and Fig \ref{fig:multi:loc_phi_2_3} show the effect of changing the locations of the peaks. We can see again that when the peaks are very sharp ($\epsilon$ very small), the spectral index diverges at the corners of the plots.  

Note that the bounds for $n_s$ given in subsection \ref{subsec:3_1} for models with $p=2/3$ and $N_e = 60$ is $[0.963, 0.978]$. The majority of the predictions in Fig \ref{fig:multi:loc_mu_2_3} and Fig \ref{fig:multi:loc_phi_2_3} are within this region. Hence except for the corner situations, the locations and number of the peaks do not affect the result significantly. If one adds more peaks in the distributions, the prediction would not change too much because adding more peaks would only make the distributions of $\hat{\mu}$ and $\hat{\phi}$ less separated and thus $n_s$ larger.

\section{Summary}
In this work we studied the possible range of the spectral index $n_s$ and the tensor-to-scalar ratio $r$ for the multifield axion monodromy models. Given the two conditions from the number of e-foldings (\ref{condition:EFolding}) and the normalization scale of the scalar curvature perturbation (\ref{condition:PR}), we showed $n_s$ is bounded above by the single field model predictions (\ref{ns:max}). Strictly speaking there is no lower bound purely by imposing the two conditions. We showed this using the explicit examples given in section \ref{sec:StrictBound} and \ref{sec:stats}.

\begin{figure}[!ht]
	\centering
	\includegraphics[width=\textwidth]{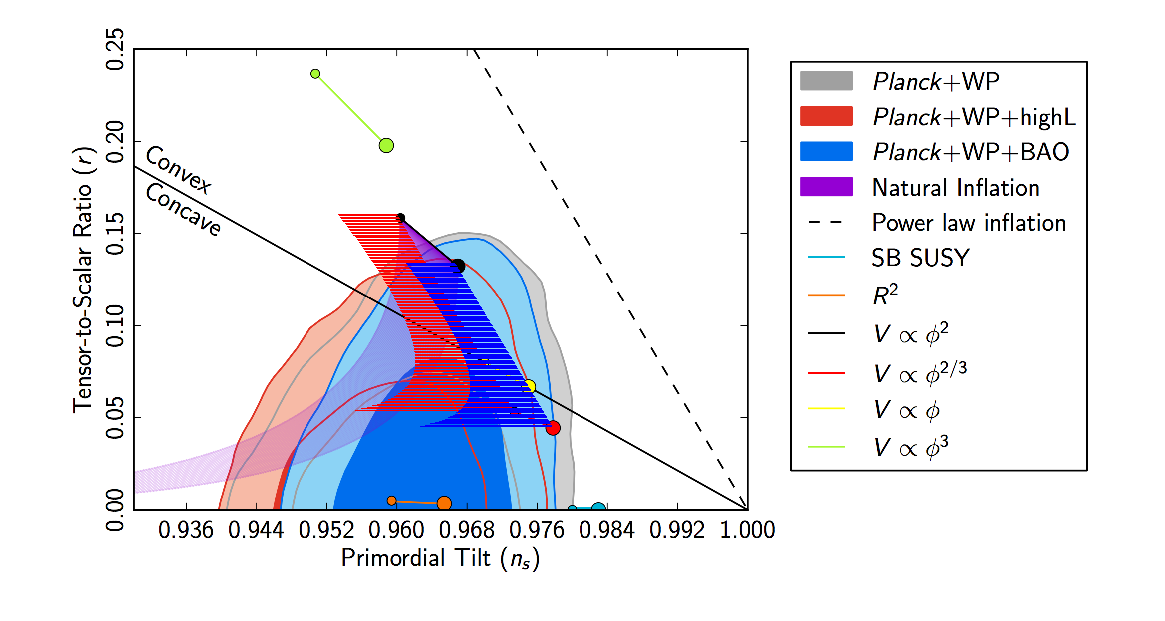}
	\caption{$r$ - $n_s$ plot for the models with $2/3 \le p \le 2$. The plot of experimental results is taken from \cite{Plack2013:I}. The blue and red shaded regions are for the cases with 60 e-foldings and 50 e-foldings, respectively. $p$ increases from the bottom to the top. We should emphasize again that it is possible to have $n_s$ below (\ref{ns:min}) if only conditions (\ref{condition:EFolding}) and (\ref{condition:PR}) are imposed, but they only constitute a very small portion of all the possible models and most of them are unphysical. Therefore those models are not included in this plot.}
	\label{fig:r-ns}
\end{figure}

For the majority of the models, however, the spectral index can be bounded below by $n_s^{\min}$ as in (\ref{ns:min}), which is given by the models with uniformly distributed ``mass'' parameters and initial conditions. Fig \ref{fig:r-ns} shows this possible region for predictions made by models with $2/3 \le p \le 2$ as overlaid with the Planck result \cite{Plack2013:I}. The spectral index is shifted to the left from the single field predictions when the models get multiple fields. Most of the predictions are well within the region given by the experimental results.

The assumptions we made here are (\ref{condition:combined}) and the decoupled power law form of the inflaton potential. The shaded regions in Fig \ref{fig:r-ns} are predicted by the majority of the models that satisfy the slow roll condition, as shown in section \ref{sec:stats}. More knowledge about the form of the potential and the initial conditions can in principle make the prediction region tighter, although among the various ways of constructing accelerated expansion solutions in string theory \cite{StringInflationReview, NewSolution}, there does not seem to be a preferred one. In the meantime, the smallness of non-gaussianity may put strong constraints on the parameters. A direct coupling between the inflaton fields may also affect the predictions. We leave these interesting topics to future work.

\section*{Acknowledgments}
I am grateful to E. Silverstein for helpful discussion and encouragement of finishing this work. This research was supported in part by the National Science Foundation under grants PHY-0756174 and NSF PHY11-25915 and by the Department of Energy under contract DE-AC03-76SF00515.


\end{document}